# Viscous electron fluids

Marco Polini[1,2] and Andre K. Geim[2]

[1]Istituto Italiano di Tecnologia, Graphene Labs, Via Morego 30, I-16163 Genova, Italy
[2]School of Physics & Astronomy, University of Manchester, Oxford Road, Manchester M13 9PL, United Kingdom

*Recent advances in materials science have made it possible to achieve conditions under which electrons in metals start behaving as highly viscous fluids, "thicker than honey", and exhibit fascinating hydrodynamic effects.*

Electrons in metals and semiconductors are often described as little balls moving around (Fig. 1), not dissimilar to atoms or molecules in dilute gases. This description is originally due to Lev Landau who managed to reduce the complex many-body problem to a simple concept of a Fermi gas of nearly-free electrons. This is particularly counterintuitive because the same Landau theory also infers that electron gases in normal metals must be atrociously viscous due to electron-electron (e-e) collisions. Their viscosity rapidly goes to infinity with decreasing temperature $T$, and simple estimates show that, at liquid helium $T$, electron gases in metals should be more viscous than, for example, honey. This leads us to a conundrum. On one hand, to describe resistivity of metals, it seems appropriate to use the hydrodynamic description with its difficult-to-solve Navier-Stokes equation. On the other hand, this is not happening apparently. Otherwise, researchers won't routinely use the ball-like model and the resulting simple Drude formula, and readers won't be taught those during their undergraduate studies.

The reason why the ball model works so well stems from the fact that a spatial range of interactions between electrons in metals is very different from that for molecules in gases. Molecules scatter at very short distances, when they come in direct contact with each other. In contrast, electrons can be very close (at distances down to interatomic) but the distance at which they effectively interact (the mean free path for e-e collisions, $\ell_{ee}$) is much longer and diverges at low temperatures. The latter leaves plenty of time and space for impurities and thermal vibrations (phonons) to destroy any nascent collective response of electrons, which would otherwise result in their viscous flow. This fact is not news but poorly appreciated even by the very people whose day job is to study the electronic properties of materials. To further this point, let us do the following *gedanken* experiment. Take a highly viscous classical gas moving through a large tube. Its flow has to be described by hydrodynamics with all its attributes. Now fill the same tube with sand so that intergranular gaps are smaller than the mean free path for the gas' molecules. The flow through the resulting porous medium is no longer viscous or Poiseuille but becomes Knudsen's. The latter is analogous to the ball model for electrons. Something similar happens in normal metals: Impurities and phonons act as those grains of sand, which are packed densely enough to eliminate any sign of electrons' collective behavior.

In theory, it should be possible to recover the "intrinsic" hydrodynamic behavior of electrons if a metal is ultraclean (to avoid impurities) and cooled down to low enough $T$ to avoid phonon scattering (consider this as removal of the sand). In practice, little experimental progress has been made in reaching the hydrodynamic regime, despite extensive efforts over many decades. Fortunately, the situation has changed recently due to the availability of new high-quality electronic materials, especially graphene.

**History of misbehaving electrons**

It was back in 1963, when Soviet theorist Radii Gurzhi posed the question about how a viscous electron flow could reveal itself in an experiment [1]. He assumed that there could exist a metallic system in which $\ell_{ee}$ provided the shortest length scale in the electron transport problem, much shorter than both sample size $W$ and the mean free path $\ell$ of electrons with respect to all collisions that did not conserve the electron momentum such as scattering on phonons, crystal defects, etc. Under this assumption, frequent collisions between electrons should be able to establish a collective flow (Fig. 1) because their total momentum (and energy) is not lost to the outside world. Gurzhi found that resistivity *R* of such an imaginary metal would have to *decrease* with increasing $T$. This is a shocking result because the standard definition of a metal system is that its *R increases* with $T$. Nonetheless, the conclusion was unambiguous and could be traced back to the fact that the electron viscosity $\nu$ in metals decreases with $T$ whereas it is obviously easier to push a fluid if it's less viscous. The found anomaly is usually referred to as the Gurzhi effect, telling us that if a metal enters the hydrodynamic regime it should exhibit a $T$ dependence opposite to metallic, more like that of a semiconductor.

Unfortunately, it turned out to be next to impossible to find a system where the required conditions $\ell_{ee} \ll W, \ell$ were met. One usually thinks of large clean crystals at cryogenic temperatures, which would mitigate the effect of lattice vibrations and thus increase $\ell$. Indeed, clean three-dimensional (3D) metals at low $T$ exhibit $\ell$ that can easily reach a good fraction of a centimeter! However, $\ell_{ee}$ also rapidly increases with decreasing $T$ because of "Pauli blocking". In other words, Fermi statistics greatly limits the available phase space for e-e collisions, if $T \ll T_F$ where $T_F$ is the Fermi temperature. As a result, $\ell_{ee}$ diverges as $(T_F/T)^2$ with decreasing $T$. This low-$T$ regime is precisely where Landau quasiparticles are long lived ($\ell_{ee} \gg W, \ell$) and the Drude, ball-like model for electrical conductivity is justified. The only way to reach the hydrodynamic regime is to work at elevated $T$ such that the Fermi sphere becomes "softer" and the Pauli blocking less obstructive for e-e scattering. Then, phonons become the main culprit, limiting $\ell$ to the electron-phonon scattering length, $\ell_{ep}$. The resulting condition $\ell_{ee} \ll \ell_{ep}$ required to observe a viscous behavior is very difficult to satisfy, especially because $\ell_{ep}$ often decreases faster with increasing $T$ than $\ell_{ee}$ (for 3D metals, $\ell_{ep}$ usually varies as $T^{-3}$, faster than $\ell_{ee} \propto T^{-2}$). This narrows materials systems and the $T$ interval where electron hydrodynamics could possibly be observed.

An elegant attempt to break the vicious cycle was undertaken in the 90s using a nonlinear transport regime [2]. Researchers applied a high electrical current to increase the electron temperature of a semiconductor 2D electron system (2DES), which shortened $\ell_{\text{ee}}$. At the same time, the crystal lattice remained close to liquid helium $T$, thus reducing electron-phonon scattering. The measured differential resistance revealed a small but clear bump as a function of applied current, which was plausibly interpreted as evidence for the Gurzhi effect. But Gurzhi and co-workers [3] immediately disagreed, pointing out that peculiarities of e-e scattering in 2D demanded even a more stringent condition $\ell_{\text{ee}} \ll W(T/T_{\text{F}})$ than that in 3D metals, which was unlikely to be achieved in the experiment. They offered an alternative explanation for the observed nonlinearity as ballistic transport affected by e-e interactions.

And this is it: For half a century after the Gurzhi theory, no electronic system was found to exhibit unambiguous signs of the hydrodynamic behavior, despite the great progress achieved in the understanding of the effect of e-e interactions on other materials properties (see Box 1). In the meantime, interest in electron hydrodynamics has continued to grow, especially because it was realized that the same physics should apply to other strongly interacting quantum systems including ultra-hot nuclear matter (quark-gluon plasmas) and ultra-cold atomic Fermi gases in the unitary limit. Paradoxically, while the viscosity of those forms of extreme quantum matter has been measured, it was not clear until recently how to probe electron hydrodynamics in solid-state devices.

**Graphene to the rescue**

Despite having the Nobel prize behind its belt, several years ago graphene did not look as a promising candidate for studies of electron hydrodynamics. It was relatively dirty with $\ell$ for impurity scattering barely exceeding 100 nm [4]. All changed around 2011 when the electronic quality of graphene was dramatically improved by encapsulating it in hexagonal boron nitride. This shielded graphene from outside impurities and also made it perfectly flat suppressing scattering at microscopic corrugations. Graphene has become one of the highest quality electronic materials ever: Its low-$T$ $\ell$ is currently limited only by the device size $W$ (up to ~10 μm) and remains longer than a micron even at room $T$. Most importantly, graphene is extremely stiff, stiffer than diamond, which suppresses phonon scattering and increases $\ell_{\text{ep}}$. Also, unlike in 3D metals, electron-phonon scattering in 2D graphene increases slowly with temperature (in graphene $\ell_{\text{ep}} \propto T^{-1}$ with a small proportionality coefficient due to the high stiffness) whereas e-e scattering grows much faster ($\ell_{\text{ee}} \propto T^{-2}$). Therefore, above a certain $T$, $\ell_{\text{ee}}$ is expected to become the smallest scattering length. In addition, graphene's $T_{\text{F}}$ is typically > 1.000 K, neither too small, as in semiconductor 2DESs where the Fermi surface is largely destroyed at room $T$, nor too high for the mentioned 2D condition $\ell_{\text{ee}} \ll W(T/T_{\text{F}})$ to have a major consequence. To sum up, it is hardly possible to imagine a better candidate than graphene to study viscous electron flows.

Against all these expectations graphene's resistivity did not show any sign of the Gurzhi effect at any $T$. Is there something wrong with all those theoretical ideas? With the benefit of hindsight, it is straightforward to understand why viscous effects did not show up previously. Indeed, the kinematic viscosity $\nu$ enters the Navier-Stokes equation as a factor in front of the second spatial derivative of the velocity $\boldsymbol{v}(x,y)$ (see Box 2). In the standard resistance measurements using a long strip of a uniform width, only the velocity component $v_x(y)$ in the flow direction $\boldsymbol{x}$ is non-zero and may depend on the transverse coordinate $y$. Unless major momentum losses occur at the strip boundaries, the dependence on $y$ tends to be weak, resulting in a fairly uniform flow profile. Without a velocity gradient, the viscosity term contributes little in the solution of the Navier-Stokes equation and, hence, in the resistance $R$. This consideration offers an interesting tip: To maximize hydrodynamics effects in experiment, it is essential to create a current flow as inhomogeneous as possible [5].

**Negative resistance and whirlpools of electrical current**

One of the geometries providing large velocity gradients is a narrow current injector (Fig. 2). In this case, the Navier-Stokes equation yields that the electric potential changes its sign at a characteristic distance of the order of $D_\nu = \sqrt{\ell_{ee}\ell}/2$ from the injector [5,6]. One can measure this local potential by placing a voltage probe very close to the current injector, using the so-called vicinity geometry (Fig. 2). The corresponding resistance $R_V$ (local voltage divided by the injected current) has the normal, positive sign for non-interacting electrons in both Drude (diffusive) and ballistic transport regimes. Negative $R_V$ provides a smoking gun for detection of a viscous flow [5]. One has to be careful however: with increasing $T$, the initial sign change of $R_V(T)$ was shown to indicate that ballistic transport is strongly affected by e-e interactions, and the hydrodynamic regime fully develops only later, at higher $T$ as e-e collisions become more frequent [7]. The observation of negative $R_V$ in graphene and comparing the behavior with the Navier-Stokes theory allowed the first measurement of electron fluid's viscosity. At liquid-nitrogen temperatures, $\nu$ turned out to be 100 times larger than that typical of honey, in quantitative agreement with the many-body theory [5].

The Navier-Stokes theory also predicts another spectacular effect in conductivity of metals due to viscosity [5,6,8]. The negative spot of electric potential near a current injector may develop into a whirlpool of electrical current (Fig. 2). This is perhaps not so surprising as whirlpools are very familiar for a laminar flow of ordinary fluids and, for example, appear around stones in a river. In the vicinity geometry [5,8], whirlpools are always expected to exist near a narrow injector and only their size $D_\nu$ depends on the actual value of $\nu$. However, for other geometries creating a non-uniform flow [6], current whirlpools generally disappear if $D_\nu$ gets smaller than the characteristic device size $W$, even though the negative potential anomaly remains well developed. Whirlpools of electrical current have yet to be observed in experiment.

**Electrons go superballistic**

In 1908, Martin Knudsen observed that a gas flow through a small aperture suddenly increased as he gradually increased the gas' density. This implies that higher viscosity boosted a gas flow, which is quite counterintuitive. The effect is of course well understood now and occurs because of the transition from the Knudsen to Poiseuille flow regime, which in the language of metal physics means from ballistic to viscous electron transport. The phenomenon observed by Knudsen can also be viewed as a preexisted analogue of the Gurzhi effect but for gas flows rather than electrons.

An experiment similar to Knudsen's was recently realized in graphene [9]. A narrow aperture of width $w$ connected two wider regions (Figs. 3 and 4), a geometry known as point contact (PC). In the ballistic regime at low-$T$, such PCs were first made and studied by Yuri Sharvin in the 60s. He found that, even in the ideal case, without any disorder and scattering, PCs exhibited a finite electrical conductance. Its value is given by the number of electron wave modes that can fit inside the aperture. Until recently, it has been tacitly accepted that Sharvin's conductance is the highest possible value. Indeed, the absence of disorder seems to imply the best case scenario for unimpeded electron transport. This is wrong. Fig. 4 shows that, upon increasing $T$ and entering the hydrodynamic regime, the resistance measured using graphene PCs dropped below the ideal ballistic limit. This drop was caused by the transition from ballistic to viscous electron transport. The resistance drop was also accompanied by a semiconductor-like $T$ dependence, which provided the first unambiguous manifestation of the long-awaited observation of the Gurzhi effect.

How is it possible that viscosity helps electrical conductivity? Indeed, we know from the basic physics that additional scattering should increase the resistance (Matthiessen's rule) whereas a viscous flow implies a lot of e-e scattering added on top of ballistic transport. The paradox allows a simple physical explanation. Crossing over from the low-$T$ regime (where Sharvin's description applies) to the hydrodynamic regime at higher $T$, electron viscosity sets a funnel-like current pattern, similarly to what happened in the Knudsen experiment. Now imagine an electron moving towards the PC (Fig. 3). In the ballistic regime, it can hit the constriction wall and then falls out of play, not contributing to the conductance. In the hydrodynamic regime, the same electron is dragged by electron collisions towards the orifice and forced to funnel through it (Fig. 3). This funneling enhances the PC conductance above Sharvin's ballistic limit. Mathematically, such a "superballistic" flow stems from an anti-Matthiessen's rule where conductivities rather than resistivities add up (Ref. [10] and Box 3). Comparison between the experimental results and theory allowed accurate measurements of graphene's $\nu$ as a function of electron concentration and $T$ (Fig. 4).

**Electronic magneto-hydrodynamics**

Another knob that can be used to explore viscous flow is the magnetic field $B$. In traditional metallic systems, $B$ is well known to cause the Hall effect, a potential drop perpendicular to directions of both current flow and magnetic field. How is the Hall effect affected by electron viscosity?

From a theory perspective, $B$ breaks down time-reversal symmetry and leads to the appearance of a new kinematic coefficient $\nu_H$ in the Navier-Stokes equation. The coefficient is called the Hall viscosity. It is odd under reversal of $B$ and dissipationless as the Lorentz force is. The Hall viscosity gives rise to an extra term in the Navier-Stokes equation, which is proportional to $\nu_H$ and acts against the Lorentz force, suppressing the resulting potential drop. This suppression of the Hall effect is very local, extending over distances of only about $D_\nu$ (Fig. 5). By placing voltage probes in the immediate vicinity of a narrow current injector, it is possible to measure a local Hall effect [11]. For graphene in the hydrodynamic regime, it was found to be notably smaller than the standard Hall effect measured simultaneously at large distances from the current contact.

**What's next**

Armed with the recently acquired knowledge of how to force electron hydrodynamics to show up in experiment, we expect viscous phenomena to be observed soon in many "old" systems including 2DESs in semiconductors, graphite, bismuth, etc. Some evidence for a viscous flow was already reported for delafossites [12], and local (vicinity and point-contact) geometries should help clarify those observations. The above also applies to strongly correlated and topological materials and, maybe, even "strange metals" such as high-temperature superconductors in the normal state, which are expected to be viscous. Materials where electrons and holes coexist and strong interactions between them must be taken into account present another interesting challenge [13,14]. Particularly enticing is to extend the existing hydrodynamic studies into the regime where nonlinear terms in the Navier-Stokes equation could no longer be ignored. For classical fluids, those terms are responsible for hard-to-understand nonlinear phenomena such as turbulence. Similar physics is expected to occur in electron fluids but the studies would require materials with smaller $\nu$ and longer $\tau$ as compared to the 2DESs studied so far.

For all the above ventures, one can obviously employ not only electrical measurements but also various visualizations tools that are available nowadays such as scanning tips with submicron sensors of either voltage or magnetic field. Those probes can image local distributions of electrical current and reveal electron hydrodynamics at the entirely new, more spectacular level. Watch out for beautiful images of electron whirlpools and viscous flows coming soon.

Box 1: **On electron-electron interactions in metallic systems**. In the 1930s, Landau and Pomeranchuk in the Soviet Union and Baber in the UK showed that the resistance of clean metals at sufficiently low $T$ should be dominated by a term proportional to $T^2$ due to (Umklapp) e-e scattering. This contribution was extensively studied in many transition and alkali metals. Interestingly, a decrease of resistance with increasing $T$ was reported for thin wires of high-purity K at $T$ below 1 K and initially attributed to the Gurzhi effect. It was later shown that the anomaly could be caused by a variety of phenomena.

In mesoscopic conductors [15], e-e interactions become essential as the dominant source of phase relaxation in both diffusive ($\ell \ll \ell_{ee}, W$) and ballistic ($W \ll \ell_{ee} \ll \ell$) regimes. A finite phase breaking length $\ell_\phi$ greatly affects all interference phenomena including weak localization, Aharonov-Bohm oscillations and universal conductance fluctuations. Those effects were widely used to measure $\ell_\phi$ but have little in common with electron hydrodynamics, the regime where $\ell_{ee} \ll \ell, W$.

Box 2: **Navier-Stokes equation in condensed matter physics**. The motion of water in oceans, turbulent air currents and "wine tears" that appear inside of a glass are only a few examples of phenomena that are governed by the Navier-Stokes equation. It is effectively Newton's second law for each "fluid element", a small volume of a liquid or gas subjected to external forces. As of today, no mathematical theory exists which would unlock the equation's full solution. Finding it remains one of the famous seven Millennium Prize Problems.

To describe a steady-state flow of electrons, the simplest, linearized form of the Navier-Stokes equation is normally used [5,6,8]
$$\frac{\sigma_0}{e}\nabla\phi(r) + D_\nu^2 \nabla^2 J(r) - J(r) = 0,$$
where $J(r) = nv(r)$ is the current density, $n$ is the electron density, $\phi(r)$ is the electric potential, $\sigma_0$ is a Drude conductivity, and $e$ is the electron charge. The length over which the flow momentum diffuses is given by $D_\nu = \sqrt{\nu\tau}$ where $\tau$ is a time scale that describes momentum dissipation due to scattering of electrons with impurities and phonons. In the limit $D_\nu \to 0$, the linearized Navier-Stokes equation yields the standard Ohm's law locally, $-eJ(r) = \sigma_0 E(r)$, where $E(r) = -\nabla\phi(r)$ is the electric field. To find an electron flow pattern, the Navier-Stokes equation needs to be solved together with the continuity equation, $\nabla \cdot J(r) = 0$, and given boundary conditions.

Box 3: **Anti-Matthiessen's rule**. Matthiessen's rule (1864) states that if there are several independent scattering processes, the total resistance $R$ is given by the sum of the resistances due to each individual process. Deviations from Matthiessen's rule do occur in metals but are generally tiny. The occurrence of "anti-Matthiessen's rule" where conductivities $G$ rather than resistivities are added, are exceptionally rare. One possible scenario was proposed for the case of "strange metals" [16].

A viscous electron flow through a point contact is another exception. There are two relevant time scales in this problem. The first one is the single-particle flight time across the constriction, $\tau_1 = \frac{2}{\pi}(w/v_F)$. The second is the time scale over which the momentum diffuses over the same distance, $\tau_2 = \frac{\pi}{32}(w^2/\nu)$. It was found [10] that the total PC resistance is given by
$$R_{PC} = \frac{1}{G_1 + G_2} = \frac{m}{n\,e^2}\frac{1}{\tau_1 + \tau_2}.$$
where $m$ is the effective electron mass. This anti-Matthiessen equation was confirmed to be valid by the experiment [9].

**Acknowledgements.**

This work was supported in part by the European Union's Horizon 2020 research and innovation programme under grant agreement No. 785219 – GrapheneCore2. We wish to thank all the people who have contributed to this research, and in particular D. A. Bandurin, A. I. Berdyugin, R. K. Kumar, F. M. D. Pellegrino, L. A. Ponomarenko, A. Principi, A. Tomadin, and I. Torre. M.P. wishes to dedicate this Article to Rachele.

**Figures**

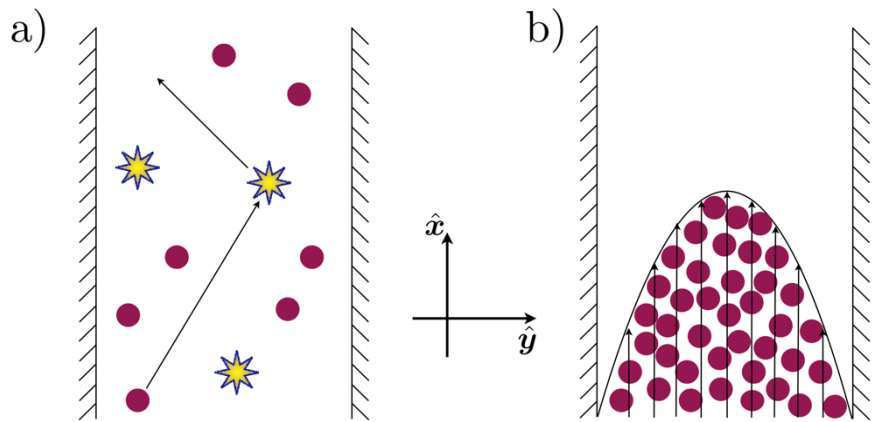

**Figure 1: Drude transport versus viscous electron flow. a,** The "ball" model. Electrons (circles) move as independent quasiparticles undergoing collisions with impurities and phonons (stars). **b,** In the hydrodynamic regime, frequent e-e interactions enable a collective state of flow dubbed "Poiseuille" flow. The velocity in the flow direction, $v_x$, is a parabolic function of the transverse coordinate $y$.

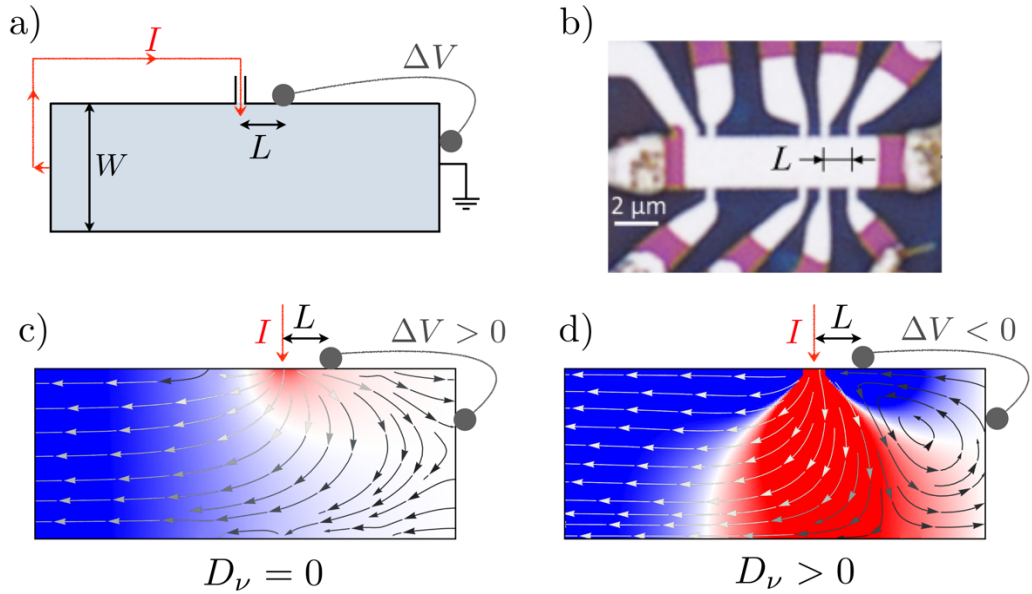

**Figure 2: The vicinity resistance and current whirlpools. a,** "Vicinity" resistance measurements [5]. Current $I$ is injected into a 2D device of width $W$ through a narrow lead, and a potential drop $\Delta V$ is measured between a voltage probe placed at a short distance $L$ from the injector and a faraway lead. The vicinity resistance is $R_V = \Delta V / I$. **b,** Optical micrograph of a multiterminal graphene device used in the experiments [5]. **c,** The color map shows the calculated distribution of electrical potential in the Ohmic regime, i.e. in the absence of viscosity ($D_\nu = 0$). The steady-state current pattern is shown by arrows. The vicinity voltage is positive. **d,** As in (c) but for the case of viscous flow ($D_\nu > 0$). Lobes of negative voltage clearly emerge near the current injector resulting in $\Delta V < 0$. The finite viscosity also induces whirlpools in the current flow [5,8].

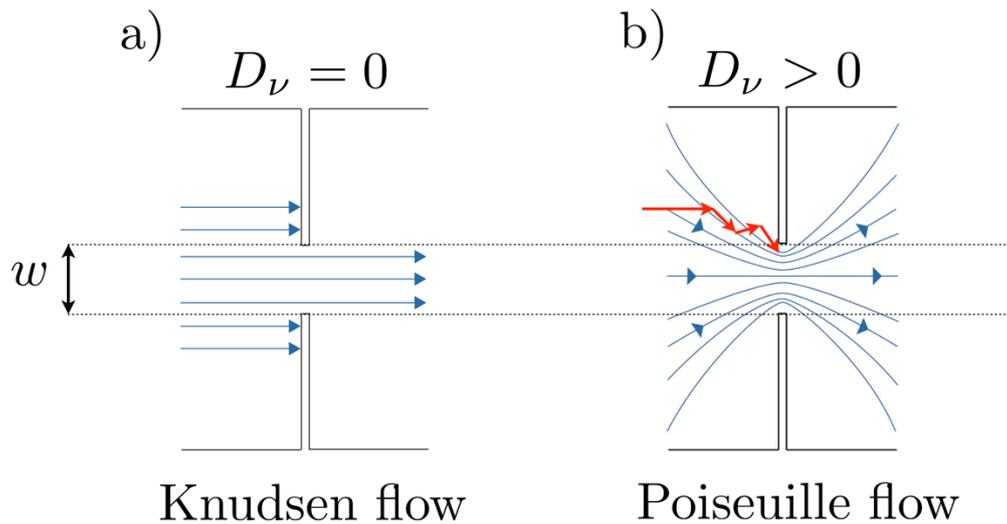

**Figure 3: Electron flow through a constriction.** A constriction of width $w \ll W$ separates two wide leads. **a,** For the case of ballistic transport, only a certain number of modes scaling like the product of the Fermi wave number $k_F$ and $w$ can fit into the constriction. Such a quantum-mechanical picture of a flow of independent electrons yields the Sharvin conductance (Box 3). **b,** For a viscous electron fluid, the Poiseuille flow corresponds to minimal momentum dissipation that occurs at the constriction boundaries. Individual electron trajectories are illustrated by red arrows: An incoming electron aiming outside the opening won't contribute to the conductance but e-e collisions "drag" such electrons towards the opening so that the resistance decreases. An "anti-Matthiessen" rule is obeyed (see Box 3).

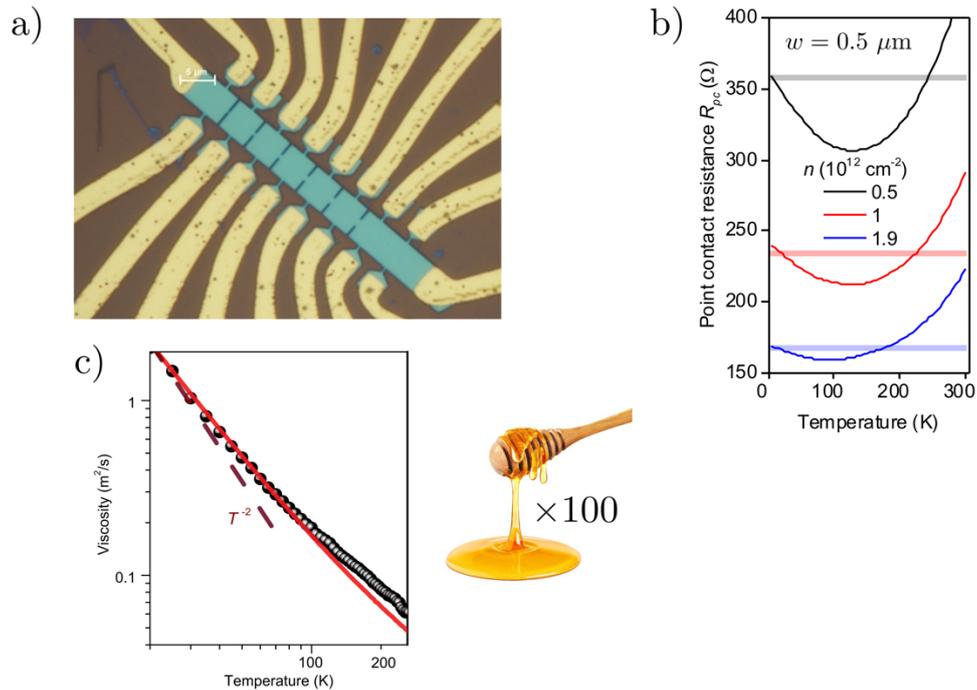

**Figure 4: Gurzhi effect in the point-contact geometry and electron viscosity. a,** Optical image of a graphene device with a series of point contacts of different widths. **b,** Resistance of one of them ($w = 0.5$ μm) as a function of temperature for three carrier densities $n$. The horizontal lines indicate the ideal, ballistic limit. As clearly seen, the resistance drops below Sharvin's value and displays a non-monotonic dependence (Gurzhi effect). **c,** Extracted viscosity $\nu$ as a function of $T$ for $n = 10^{12} \text{cm}^{-2}$ [9]. The experiment (circles) closely agree with many-body theory calculations (red curve). Note the y-axis scale: Typical viscosity of honey is $\approx 10^{-3}$ m$^2$/s.

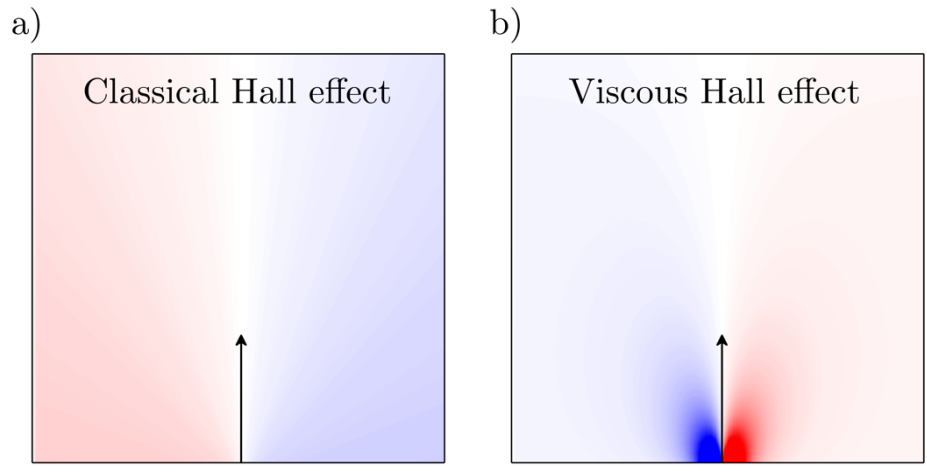

**Figure 5: Classical versus viscous Hall effect. a,** Spatial distribution of the electrical potential near a narrow current injector (black arrow) in a 2D electron system subjected to a perpendicular magnetic field (red: positive; blue negative). **b,** Same as in (a) but for the case of a highly viscous electron flow [11]. If compared with (a), the potential has the opposite sign and is localized at a distance of about $D_v$ near the current injector.